\documentclass[aps,prd,notitlepage,superscriptaddress,longbibliography,nofootinbib]{revtex4-2}
\usepackage[dvipdfmx]{graphicx}

\usepackage{dcolumn}    
\usepackage{bm}         
\usepackage{hyperref}   
\usepackage{bmpsize}
\usepackage[english]{babel}
\usepackage[skip=1pt, justification=raggedright]{caption}
\usepackage{graphicx}
\vspace{-10pt}
\usepackage{amsmath}
\hypersetup{
    colorlinks=true,
    citecolor=blue,
    linkcolor=blue,
    urlcolor=blue
}

\begin{document}

\title{CHRONOS: Cryogenic sub-Hz cROss torsion bar detector with quantum NOn-demolition Speed meter}

\author{Yuki Inoue}
\thanks{Corresponding author: iyuki@ncu.edu.tw}
\affiliation{Department of Physics,National Central University, Taoyuan, Taiwan}
\affiliation{Center for High Energy and High Field (CHiP), National Central University,  Taoyuan, Taiwan}
\affiliation{Institute of Physics, Academia Sinica, Taipei, Taiwan}
\affiliation{Institute of Particle and Nuclear Studies, High Energy Acceleration Research Organization (KEK), Tsukuba, Japan}

\author{Hsiang-Chieh Hsu}
\affiliation{Institute of Physics, Academia Sinica, Taipei, Taiwan}

\author{Hsiang-Yu Huang}
\affiliation{Department of Physics,National Central University, Taoyuan, Taiwan}
\affiliation{Center for High Energy and High Field (CHiP), National Central University,  Taoyuan, Taiwan}

\author{M.Afif Ismail}
\affiliation{Department of Physics,National Central University, Taoyuan, Taiwan}
\affiliation{Center for High Energy and High Field (CHiP), National Central University,  Taoyuan, Taiwan}
\affiliation{Institute of Physics, Academia Sinica, Taipei, Taiwan}

\author{Vivek Kumar}
\affiliation{Institute of Physics, Academia Sinica, Taipei, Taiwan}
\affiliation{Department of Physics, Institute of Applied Sciences and Humanities, GLA University, Mathura 281406, India.}

\author{Miftahul Ma'arif}
\affiliation{Department of Physics,National Central University, Taoyuan, Taiwan}
\affiliation{Center for High Energy and High Field (CHiP), National Central University,  Taoyuan, Taiwan}

\author{Avani Patel}
\affiliation{Department of Physics,National Central University, Taoyuan, Taiwan}
\affiliation{Center for High Energy and High Field (CHiP), National Central University,  Taoyuan, Taiwan}

\author{Daiki Tanabe}
\affiliation{Institute of Physics, Academia Sinica, Taipei, Taiwan}
\affiliation{Center for High Energy and High Field (CHiP), National Central University,  Taoyuan, Taiwan}
\affiliation{Institute of Particle and Nuclear Studies, High Energy Acceleration Research Organization (KEK), Tsukuba, Japan}

\author{Henry Tsz-King Wong}
\affiliation{Institute of Physics, Academia Sinica, Taipei, Taiwan}
\affiliation{Center for High Energy and High Field (CHiP), National Central University,  Taoyuan, Taiwan}

\author{Ta-Chun Yu}
\affiliation{Department of Physics,National Central University, Taoyuan, Taiwan}
\affiliation{Center for High Energy and High Field (CHiP), National Central University,  Taoyuan, Taiwan}
\date{\today}

\begin{abstract}
We propose a next-generation ground-based gravitational-wave detector, Cryogenic sub-Hz cROss torsion-bar detector with quantum NOn-demolition Speed meter (CHRONOS), 
optimized for the unexplored $0.1$--$10\,\mathrm{Hz}$ band between the 
space-based LISA and future ground-based detectors such as Cosmic Explorer and the Einstein Telescope. 
CHRONOS combines a ring-cavity Sagnac interferometer with torsion-bar test masses to realize the first quantum nondemolition (QND) measurement of angular momentum in a macroscopic system. 
By implementing a speed-meter readout in the rotational degree of freedom, CHRONOS coherently cancels quantum radiation-pressure noise and enables sub-Hz observations. 
We calculate, for the first time, that detuned power-recycling and cavity-length optimization can simultaneously relax technical requirements on both torsion bars and speed meters. 
Assuming a realistic optical design with 1m torsion bar, we estimate strain sensitivities of $h \simeq 5\times10^{-19}\,\mathrm{Hz^{-1/2}}$ at $2\,\mathrm{Hz}$ for detectors with arm lengths of $2.5$ m, $40$ m, and $300$ m. 
These sensitivities enable (i) direct detection of intermediate-mass black hole binaries up to 340\,Mpc with SNR=3, 
(ii) probing SGWB down to $\Omega_{\mathrm{GW}}\sim\ 3\times10^{-4}$ at 0.2 Hz with 5 year accumulation. 
Furthermore, CHRONOS enable to prompt detection of gravity-gradient signals from M 5.5 earthquakes even with a $2.5$ m prototype. 
CHRONOS thus opens new opportunities for quantum-limited geophysical observation and multi-band, multi-messenger gravitational-wave astronomy.
\end{abstract}

\maketitle
\section{Introduction}
The first direct detection of gravitational waves (GWs) by LIGO~\cite{LIGO2016} opened the era of GW astronomy, followed by Virgo~\cite{Virgo2015} and KAGRA~\cite{KAGRA2021}, which operate above $10\,\mathrm{Hz}$ and have provided key insights into compact binaries and fundamental physics. 
At much lower frequencies, space-based missions such as LISA~\cite{LISA2017} will explore the millihertz band ($10^{-4}$--$10^{-1}\,\mathrm{Hz}$), targeting supermassive black hole mergers and cosmological signals.

However, the intermediate range of $\sim0.1$--$10\,\mathrm{Hz}$ remains largely unexplored despite its rich science potential, including intermediate-mass black hole (IMBH) binaries, stochastic gravitational-wave backgrounds (SGWBs), and prompt gravity-gradient signals from seismic events. 
Bridging this sensitivity gap requires fundamentally new detection principles.

Torsion-bar-based detectors such as TOBA~\cite{Ando2010} and TorPeDO~\cite{TorPeDO2019} demonstrated sub-Hz feasibility but the fundamental quantum back-action limits in displacement readout, requiring massive test masses for ensuring sensitivity below $1\,\mathrm{Hz}$ ~\cite{Ando2010}. 

We propose Cryogenic sub-Hz cROss torsion-bar detector with quantum NOn-demolition Speed meter (CHRONOS), a next-generation detector combining a triangular Sagnac interferometer with torsion-bar test masses to realize the first quantum nondemolition (QND) measurement of angular momentum in a macroscopic system~\cite{Caves1980RMP,Braginsky1980Science}. 
QND is a system that can break the trade-off between shot noise and back-action noise arising from quantum commutation relations, and it has been reported that this can be realized through squeezing techniques~\cite{Braginsky1996,Danilishin2012} and speed meters~\cite{Braginsky:1990ei,Purdue2002,Khalili2002}.

A novel speed-meter readout cancels radiation-pressure noise, and power-recycling mirror detuning further suppresses the low-frequency noise with coating losses.
These issues represent technical challenges for enhancing the sensitivity of torsion bars and for the first demonstration of a speed meter, and addressing them with a speed meter and a dual-recycled torsion-bar system is a unique approach.

In this paper, we explicitly evaluate the sensitivity for different test facility scales: a 2.5 m arm length, corresponding to a Glasgow-type 10 m interferometer~\cite{Robertson1995Glasgow10m}; a 40 m arm length~\cite{Chase:40mCharacterization}, corresponding to the Caltech 40 m interferometer; and a 300 m arm length, corresponding to the TAMA300~\cite{Ando:TAMA300Stable, Tatsumi:TAMA300dev,LIGO_TAMA_BurstUL2005}.
\section{Detector Concept}
The CHRONOS detector introduces a new paradigm for low-frequency GW detection by combining torsion-bar test masses, triangular ring cavities, and a dual-recycled Sagnac interferometer. 
Torsion-based detectors such as TOBA~\cite{Ando2010} and TorPeDO~\cite{TorPeDO2019} have demonstrated sub-Hz angular displacement measurements $\theta$ by exploiting correlated torsion-bar motion. 
In these systems, the response of a torsion bar in the $x$--$y$ plane with moment of inertia $I=\int \rho(x^2+y^2)dV$
 is obtained as
$\theta(\Omega) = \eta \,(h_{+}F_{+} + h_{\times}F_{\times}),$
where GW polarizations $h_{+,\times}$, antenna patterns $F_{+,\times}$, and geometrical factor $\eta = I/\int\rho(x^2-y^2)dV$ are assumed. The test-mass density $\rho$  and volume $V$ are assumed. The signal sideband frequency is obtained with $\Omega$ . 
For all cases, we assume identical torsion-bar geometries with length of torsion bar $L_{bar}=1~\mathrm{m} $, moment of inertia $I = 19.9~ \mathrm{kg~m^2}$, test-mass mass $M = 171 ~\mathrm{kg}$, test-mass temperature $T=10~\mathrm{K}$, and a geometrical factor $\eta = 0.936$. 
Assuming the torsional resonance lies well below the observation band, these schemes are fundamentally limited by quantum radiation-pressure back-action noise below $1\,\mathrm{Hz}$.

\begin{figure}[htbp]
    \centering
    \includegraphics[width=1 \linewidth]{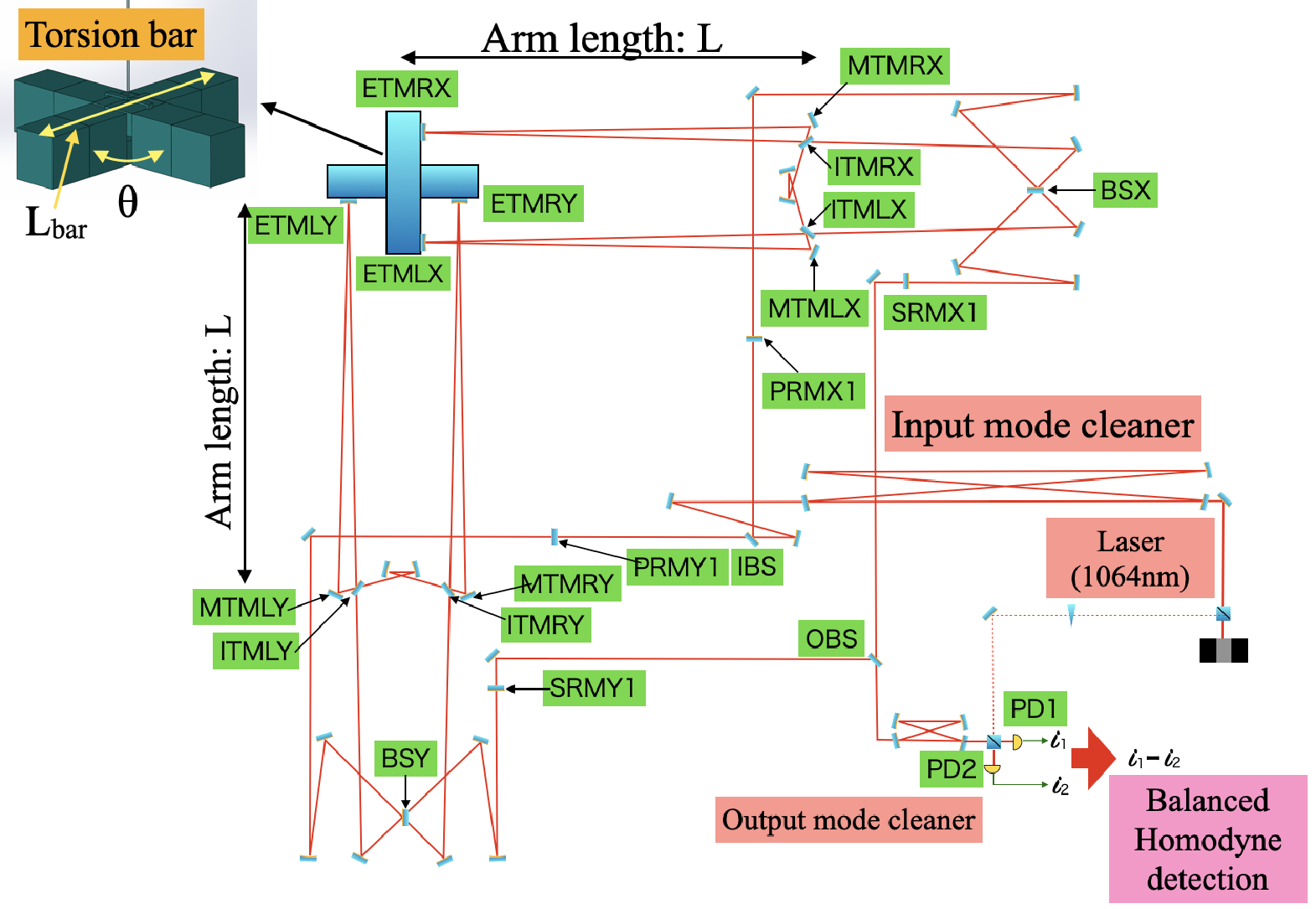}
    \caption{
        Optical configuration of CHRONOS. 
        The Sagnac interferometer integrates:
        (i) torsion-bar test masses at the center,
        (ii) Ring cavity with Sagnac interferometer at both arm,
        (iii) dual-recycled cavities with SRM and PRM.
    }
    \label{fig:optical_config}
\end{figure}

Figure~\ref{fig:optical_config} shows the CHRONOS optical configuration. 
Two orthogonal torsion bars are suspended in a cross geometry, each hosting a triangular ring-cavity. 
After pre-isolation, the input beam passes through an input mode cleaner and is split at the input beam splitter (IBS). 
The beams enter the $X$ and $Y$ arms, each forming a dual-recycled speed-meter cavity with a power-recycling mirror (PRM) as PRMY1and PRMX1, and signal-recycling mirror (SRM) as SRMY1and SRMX1. 
Outputs are recombined at the output beam splitter (OBS), filtered by an output mode cleaner, and read out via balanced homodyne detection. 
The detailed optical design of CHRONOS is described in Inoue $\mathit{et~al.}$~\cite{Inoue:inprep}.
As the $X$ and $Y$ arms are identical, we focus on the $Y$ arm below.

CHRONOS overcomes the limitations of conventional differential angle readout by directly measuring the conjugate angular momentum \(L\). This enables, for the first time, a QND measurement of \(L\) in a macroscopic system with \(O(100\,\mathrm{kg})\) test masses.

As shown in Fig.~\ref{fig:optical_config}, counter-propagating beams circulate in a triangular cavity formed by ETMLY--MTMLY--ITMLY and ETMRY--MTMRY--ITMRY, and are separated at BSY. The optical path delay between the two beams provides direct access to the angular momentum through $L(t) \propto \frac{\theta(t+\tau)-\theta(t)}{\tau}$,
where \(\tau\) denotes the optical delay time.

In a conventional differential angle readout, the time derivative of the angle satisfies
\(\dot{\theta} \propto [\hat{\theta},\hat{H}] \neq 0\),
implying that the future evolution of \(\theta\) is disturbed by measurement back-action.
In contrast, the angular momentum satisfies
\(\dot{L} \propto [\hat{L},\hat{H}] = 0\),
ensuring that \(L\) is unaffected by the measurement.
Consequently, $[\hat{L}(t),\hat{L}(t')] = 0 $,
demonstrating that \(L\) is a QND observable.

QND is a system that can break the trade-off between shot noise and back-action noise arising from quantum commutation relations, and it has been reported that this can be realized through squeezing techniques~\cite{Caves1980RMP,Braginsky1996,Danilishin2012} and speed meters~\cite{Braginsky1980Science,Purdue2002,Khalili2002}. By circumventing the quantum trade-off between shot noise and radiation-pressure noise~\cite{Ozawa2003,Ozawa2004}, these approaches provide a unique pathway toward surpassing the conventional quantum limits in GW detectors.
\section{Noise Budget and Sensitivity}
\label{sec:noise}
We evaluate the sensitivity of CHRONOS using both an analytical quantum-operator approach and full simulations performed with \textsc{Finesse3}~\cite{Brown2021}. 
The unique aspects of this study are the implementation of power recycling cavity (PRC) detuning and the integration of a speed-meter topology with torsion-bar test masses. 
Balanced homodyne detection is assumed throughout this study.~\cite{Adhikari2014}.
\begin{figure*}[t]
  \centering
  \includegraphics[width=\textwidth]{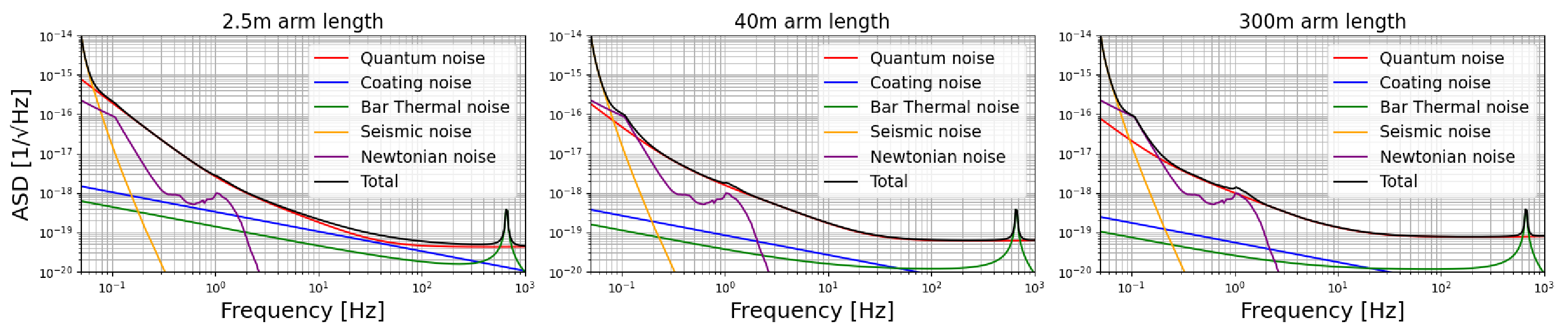} 
  \caption{Predicted CHRONOS strain sensitivity (black) and noise contributions for 2.5m, 40m and 300m cases. }
  \label{fig:noise_budget}
\end{figure*}
Following the Buonanno-Chen formalism~\cite{BuonannoChen2001,BuonannoChen2002,Chen2003SagnacSRC,Yamamoto2009MSI,Purdue2002QND}, we use the two-photon representation~\cite{Caves1980RMP}.
In this method, we define the input and output quadrature vectors as $\boldsymbol{a} = (a_1 ~a_2)^t$ and $\boldsymbol{b} = (b_1 ~ b_2)^t$,
where the input quadrature vector of the interferometer $\boldsymbol{a}$ is denoted.
Following the generalized framework of Buonanno and Chen incorporating detuning terms for both the power-recycling cavity (PRC) and the signal-recycling cavity (SRC), 
\begin{equation}
\boldsymbol{b}
= \frac{1}{M}\biggl[
  e^{2i(\beta_{\mathrm{sag}}+\Phi_s)} C \boldsymbol{a}
  + \sqrt{2\mathcal{K}_{\mathrm{sag}}}\, t_s e^{i(\beta_{\mathrm{sag}}+\Phi_s)}
  \boldsymbol{D}\frac{\theta}{\theta_{\mathrm{SQL}}}
\biggr],
\label{eq:IOrelation}
\end{equation}
where time delay of SR cavity is defined as $\Phi_s$, and the effective Sagnac phase shift is $\beta_{sag} = 2 \beta  + \pi/2$~\cite{Chen2003}
the angular displacement noise can be expressed as
\begin{equation}
    \theta =
    \frac{\theta_{\mathrm{SQL}}}{\sqrt{2(1-r_s^2)\kappa_{\mathrm{sag}}}}
    \sqrt{\frac{\boldsymbol{\nu}^{\mathsf{T}} G \boldsymbol{\nu}}{\boldsymbol{\nu}^{\mathsf{T}} Q \boldsymbol{\nu}}} ,
\end{equation}
where the standard quantum limit (SQL) of angular displacement noise ~\cite{Ando2010} is calculated as
$\theta_{\mathrm{SQL}} = \sqrt{2\hbar / (I \Omega^2)}.$
The homodyne detection vector $\boldsymbol{\nu}^{\mathsf{T}} = (\cos\zeta,\,\sin\zeta)$ is defined by the  homodyne angle $\zeta$.  
 The amplitude reflectivity of the SRM is defined as $r_s$ . We define $G = C C^{\mathsf{T}}$, and  $Q = \Re(\boldsymbol{D})\,\Re(\boldsymbol{D})^{\mathsf{T}} + \Im(\boldsymbol{D})\,\Im(\boldsymbol{D})^{\mathsf{T}}$,
where the matrix $C = [c_{ij}]$ and the column vector $\boldsymbol{D}^{\mathsf{T}}=(D_1, D_2)$ are defined following Eqs. (2.22)-(2.24) of Ref. ~\cite{BuonannoChen2001}. 

The optomechanical coupling factor $\kappa$ defined in ~\cite{BuonannoChen2001,Chen2003SagnacSRC,DanilishinKhalili2012,Graef2014ProofOfPrinciple} is modified as 
$\kappa_{\mathrm{sag}} = 4 \kappa\, |H_{\mathrm{PRC}}|^2 \sin^2\beta_{\mathrm{sag}}$, 
where transfer function of PRC is
\begin{equation}
    H_{\mathrm{PRC}} = \frac{t_p}{1-i r_p |r_{\mathrm{ifo}}|\sin{\beta_{sag}}e^{2i (\phi_p+\tau_p \Omega)}}, \label{eq:H_PRC}.
\end{equation}
 The amplitude transmission $t_p$ and reflection coefficients $r_p$ of the PRC mirror are assumed.
$r_{ifo}$ is the effective complex reflectivity of the interferometer, including SRC detuning and internal losses.
 PRC detuning phase and one-way propagation time are obtained with $\phi_p$ and $\tau_p$.
The above expression can be diagonalized along the principal axes:
\begin{equation}
    \theta =
    \frac{\theta_{\mathrm{SQL}}}{\sqrt{2\,\Gamma(\phi_s,\zeta)\,\kappa_{\mathrm{sag}}}}
    \sqrt{1 + \big(\cot\zeta_{\mathrm{eff}} - \kappa_{\mathrm{eff}}\big)^2} ,
\end{equation}
where $ \Gamma(\phi_s,\zeta)
    = (1-r_s^2)\,
    \sqrt{|D_1|^2 |D_2|^2 - \Re(D_1 D_2^{*})^2}$,
and the effective optomechanical coupling derived from $G$ and $Q$ is
\begin{equation}
    \kappa_{\mathrm{eff}}
    = \sqrt{
    \frac{
    Q_{22}G_{11} + Q_{11}G_{22} - 2Q_{12}G_{12}
    }{
    \sqrt{\det Q}\,\sqrt{\det G}
    } - 2 } .
\end{equation}
The effective homodyne angle is given by $\zeta_{\mathrm{eff}} = \zeta - \psi(f) + \theta(f)$ ,
where $\psi(f)$ and $\theta(f)$ denote the principal-axis rotations of $G$ and $Q$, respectively.
We evaluate the strain sensitivity for three detector scales optimized for the sub-Hz regime, with the design parameters chosen such that the sensitivity peaks around $1,\mathrm{Hz}$. A systematic optimization study of the interferometer parameters—specifically the detuning angle, homodyne angle, and optical coating properties, optimization study for each parameter, and definition of each noise—has been carried out in  Inoue~\textit{et al.}  and Tanabe~\textit{et al.}~\cite{Inoue:inprep,Tanabe:inprep}.
Seismic noise and Newtonian noise are estimated assuming an underground environment, based on measured ground motion data at the Kamioka site~\cite{Michimura_2017}. For the Newtonian noise, we assume that the dominant contribution arises from Rayleigh waves propagating along the ground surface.
Figure~\ref{fig:noise_budget} shows the resulting sensitivity curves for the three detector scales considered in this study, corresponding to arm lengths of 2.5 m, 40 m, and 300 m. In all cases, the interferometer parameters are optimized to target sub-Hz observations, with the peak sensitivity tuned to approximately $1,\mathrm{Hz}$. Each panel displays the individual noise contributions—quantum noise, coating thermal noise, bar thermal noise, seismic noise, and Newtonian noise—together with the total noise obtained by summing these components.

For the 2.5 m arm-length configuration, the sensitivity at low frequencies is dominated by seismic noise, while quantum noise limits the performance in the frequency range of approximately $0.1$--$1,\mathrm{Hz}$. At frequencies above $1,\mathrm{Hz}$, coating thermal noise  become the primary limiting factors.
Extending the arm length to 40 m significantly reduces both quantum noise and bar thermal noise, leading to a substantial improvement in sensitivity over a range from $0.3$ to $10,\mathrm{Hz}$. Although Newtonian noise remains an important limitation at the lowest frequencies, noise contributions originating from the detector itself are considerably suppressed at this scale.

For the 300 m arm-length configuration, quantum noise and bar thermal noise are further suppressed across the sub-Hz to few-Hz band.  This result highlights that, while scaling up the detector size is highly effective in reducing internal noise sources in torsion-bar GW detectors, environmental gravity noise ultimately emerges as the dominant sensitivity limitation.
Overall, these results demonstrate that the combination of QND techniques and a speed-meter configuration in CHRONOS, together with detector scaling, enables effective suppression of quantum noise and internal thermal noise in the sub-Hz band. Achieving further sensitivity improvements will require advances in Newtonian noise mitigation strategies as well as the development of lower-loss optical coating technologies.
\begin{table}[t]
  \caption{Parameters for sensitivity estimates.
  Definitions: $R_i=r_i^2$, $R_p=r_p^2$, $R_s=r_s^2$;
  $P_{\mathrm{in}}$ input power, $P_{\mathrm{arm}}$ arm power;
  $\phi_p$ PRC detuning, $\phi_s$ SRC detuning, $\zeta$ homodyne angle;
  $F_{\mathrm{ring}}$ finesse, $w$ beam radius at ETM.}
  \label{tab:parameters}
  \begin{tabular}{c|c|c|c}
    & 2.5 m & 40 m & 300 m \\
    \hline
    $R_s, R_p, R_i$                 & 0.5, 0.9, 0.9999      & 0.5, 0.95, 0.999      & 0.5, 0.99, 0.995 \\
    $P_{\mathrm{in}}, P_{\mathrm{arm}}$ & 1 W, 444 W             & 20 W, 2391 W          & 100 W, 18.3 kW   \\
    $\phi_p, \phi_s, \zeta$         & $-85^\circ, 0^\circ, 46^\circ$
                                    & $26^\circ, 0^\circ, -50^\circ$
                                    & $41^\circ, 2^\circ, -66^\circ$ \\
    $F_{\mathrm{ring}}$             & $3.14\times 10^{4}$    & $3.14\times 10^{3}$   & $6.27\times 10^{2}$ \\
    $w$ at ETM                      & 2 mm                   & 20 mm                 & 35 mm \\
  \end{tabular}
\end{table}
Extending the arm length shifts the radiation-pressure noise crossover frequency toward lower frequencies, enhancing low-frequency performance. Assuming a coating loss of $0.1\,$ppm quantum noise is strongly suppressed compared to conventional designs; in an ideal lossless system, the suppression is even more significant. Increasing the arm length also improves the effective optical delay  while reducing the required finesse, thereby relaxing coating-loss requirements.

A key feature of CHRONOS is its use of PRC detuning in Eq.(\ref{eq:H_PRC}), which differs from the conventional approach of detuning the SRM. Focusing on the PRC transfer function, the detuning angle of the PRC acts simultaneously on both the clockwise (CW) and counter-clockwise (CCW) components of the Sagnac speed meter~\cite{Chen2003}. This phase shift propagates into the effective homodyne angle, $\zeta_{\mathrm{eff}}$ as a term $\arg(H_{\mathrm{prc}})$, so by controlling the frequency-dependent phase balance of the CW and CCW beams through the PRC, one can effectively impart a frequency-dependent homodyne angle. In addition to enhancing the carrier field and increasing circulating power, PRC detuning enables sensitivity shaping with engineered frequency dependence, thereby suppressing radiation-pressure noise and reducing shot noise.

\section{Scientific Reach}
\begin{figure}[htbp]
    \centering
    \includegraphics[width=0.95\linewidth]{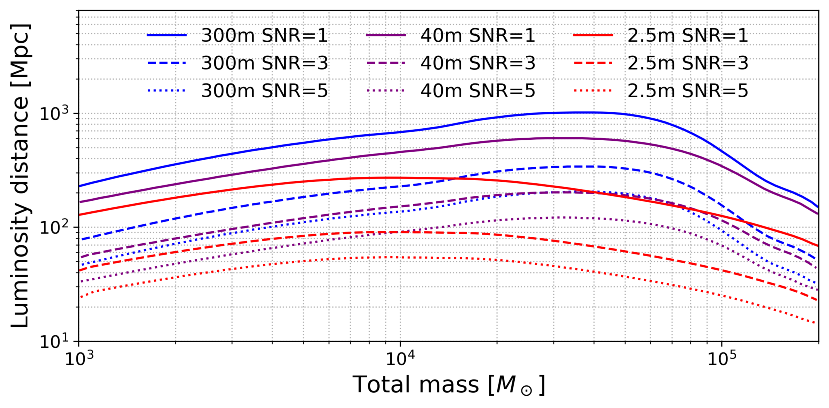}
    \caption{
        Estimated SNR for IMBH binary mergers with CHRONOS. 
    }
    \label{fig:IMBH}
\end{figure}

Three science packages are discussed for the scientific outcomes assuming each sensitivity as follows.
First, CHRONOS targets IMBH binaries with component masses of $1\times10^2$--$5\times10^5\,M_\odot$~\cite{LISA2017,Sesana2016,Miller2004,AmaroSeoane2017}. 
These signals enter the CHRONOS band weeks to days before coalescence, enabling multi-band observations in synergy with LISA and LIGO~\cite{LISA2017,LIGO2016}. 
For $M\sim4\times10^4\,M_\odot$, CHRONOS can detect mergers up to 340\,Mpc at ${\rm SNR}=3$, allowing the first statistical studies of IMBH populations and insights into supermassive black hole formation as shown in Fig.~\ref{fig:IMBH}. 
Moreover, at low frequencies, the use of high-finesse cavities causes coating losses that can destroy the QND system. This effect corresponds to the achievable finesse and the time delay $\tau$, but by increasing the arm length, an equivalent time delay can be realized with fewer reflections. As a result, low-frequency QND measurements can be maintained, leading to an increased signal to noise ratio (SNR) for high-mass IMBH binaries.
\label{sec:science}

\begin{figure}[htbp]
    \centering
    \includegraphics[width=0.95\linewidth]{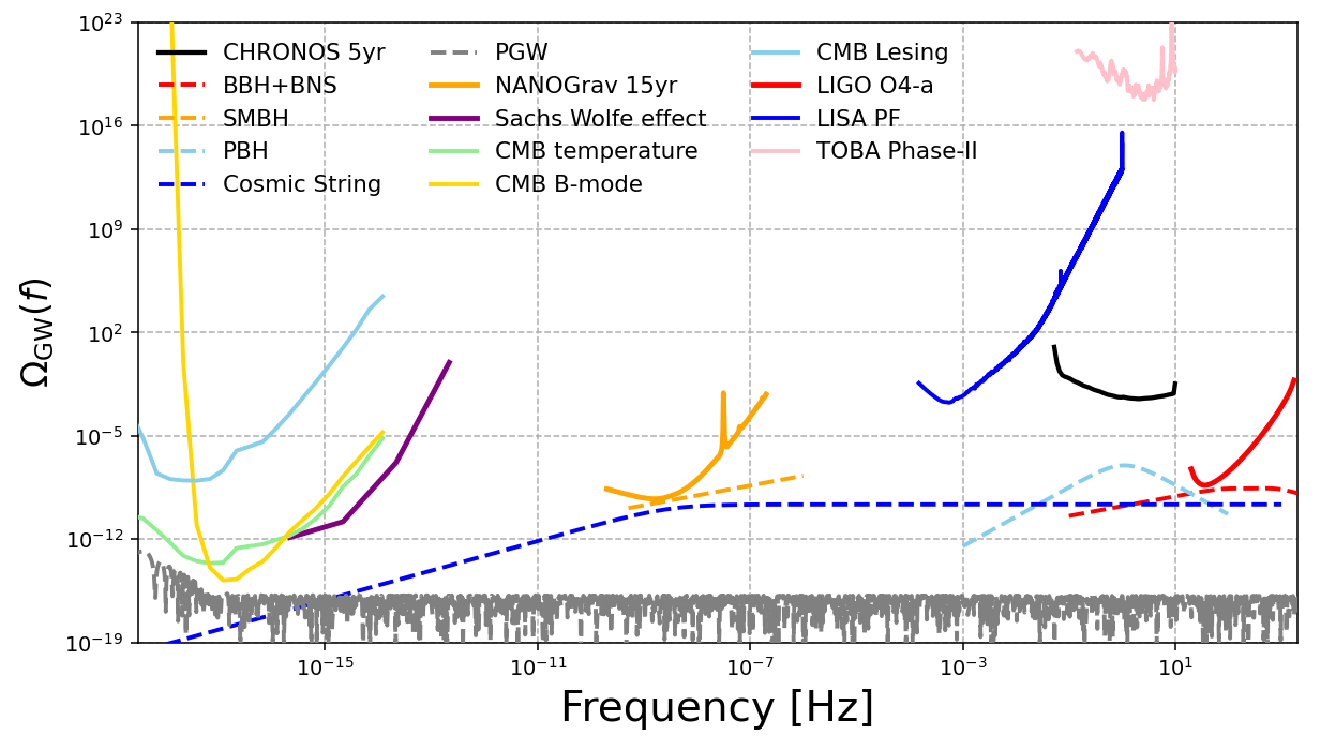}
    \caption{
        Expected sensitivity of CHRONOS to the SGWB.
LIGO O3~\cite{Abbott:2021xxi}, LISA pathfinder~\cite{Boileau:2022figures},Sachs Wolfe effect~\cite{Ng:2022prdRedshiftSGWB},CMB B-mode($r<0.06$), CMB Lensing, CMB Temperature~\cite{Namikawa:2019tax}, show the current observational limit. Binary blackhole and Binary Neutron star (BBH+BNS)~\cite{Abbott2019IsotropicStochasticO2}, Super massive blackhole (SMBH)~\cite{Campeti2021Measuring}, Primordial Blackhole (PBH)~\cite{Kohri:2018awv,Domenech:2021ztg}, Nambu-Goto type cosmic string ($G\mu\sim 10^{-8}$, $\sigma=0.1$)~\cite{Siemens:2006vk,Blanco-Pillado:2017oxo,Auclair:2019wcv}, with  Primordial Gravitational Wave (PGW)~\cite{Campeti2021Measuring} are theoretical predictions.}
    \label{fig:SGWB}
\end{figure}

Second, CHRONOS bridges the sensitivity gap between LISA~\cite{LISA2017} and ground-based detectors~\cite{LIGO2016}, enabling constraints on the SGWB in the $0.1$--$10\,\mathrm{Hz}$ band. 
Astrophysical foregrounds include unresolved black hole and neutron star binaries, while cosmological sources may arise from first-order phase transitions, cosmic strings, primordial black holes, and inflationary relics~\cite{Christensen2019}. 
Figure~\ref{fig:SGWB} shows the predicted sensitivities by assuming five year of integration. 
CHRONOS achieves an energy-density sensitivity of $\Omega_{\mathrm{GW}}\sim 3\times10^{-4}$ at 0.2 Hz, enabling new constraints on the search of new SGWB, such as primordial black holes.

\begin{figure}[htbp] 
    \centering
    \includegraphics[width=0.95\linewidth]{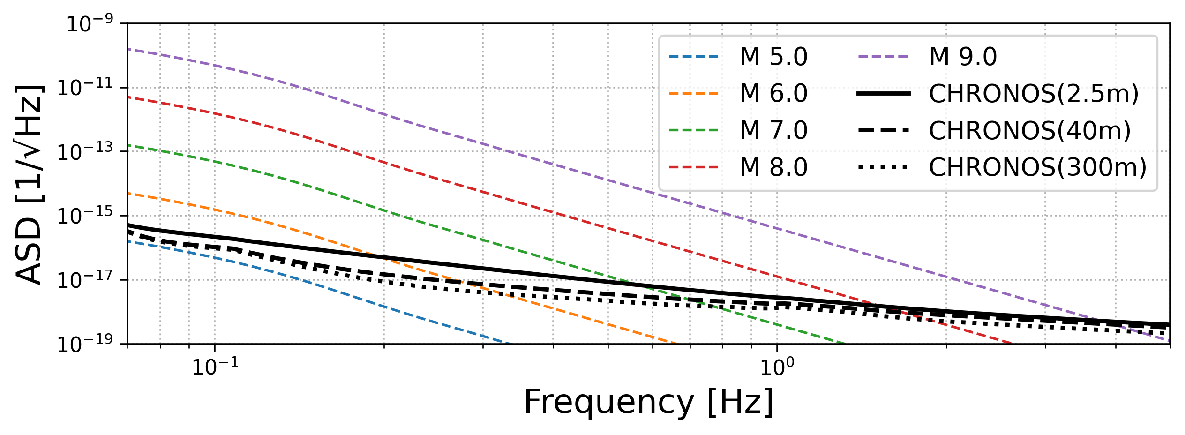}
    \caption{
        Predicted sensitivity to prompt gravity signals from large earthquakes.
        By detecting gravitational perturbations at the speed of light,
        CHRONOS enables warnings before surface-wave arrival.
    }
    \label{fig:EQ}
\end{figure}

Finally, CHRONOS enables detection of prompt gravity-gradient signals from large earthquakes as shown in Fig~\ref{fig:EQ}. 
These signals arise from rapid mass redistribution and propagate at the speed of light, potentially allowing detection seconds before destructive surface waves~\cite{Harms2015,Montagner2016,Vallee2017}. 
Simulations for the ASGRAF~\cite{ASGRAF} site near Taipei show high detection efficiency within a $100~\mathrm{km}$ radius. Taking into account the density of the ground, the minimum duration time of the arrived surface wave is estimated to be 8.3 sec. 
\section{Discussion}
\begin{figure}[htbp]
    \centering
    \includegraphics[width=0.95\linewidth]{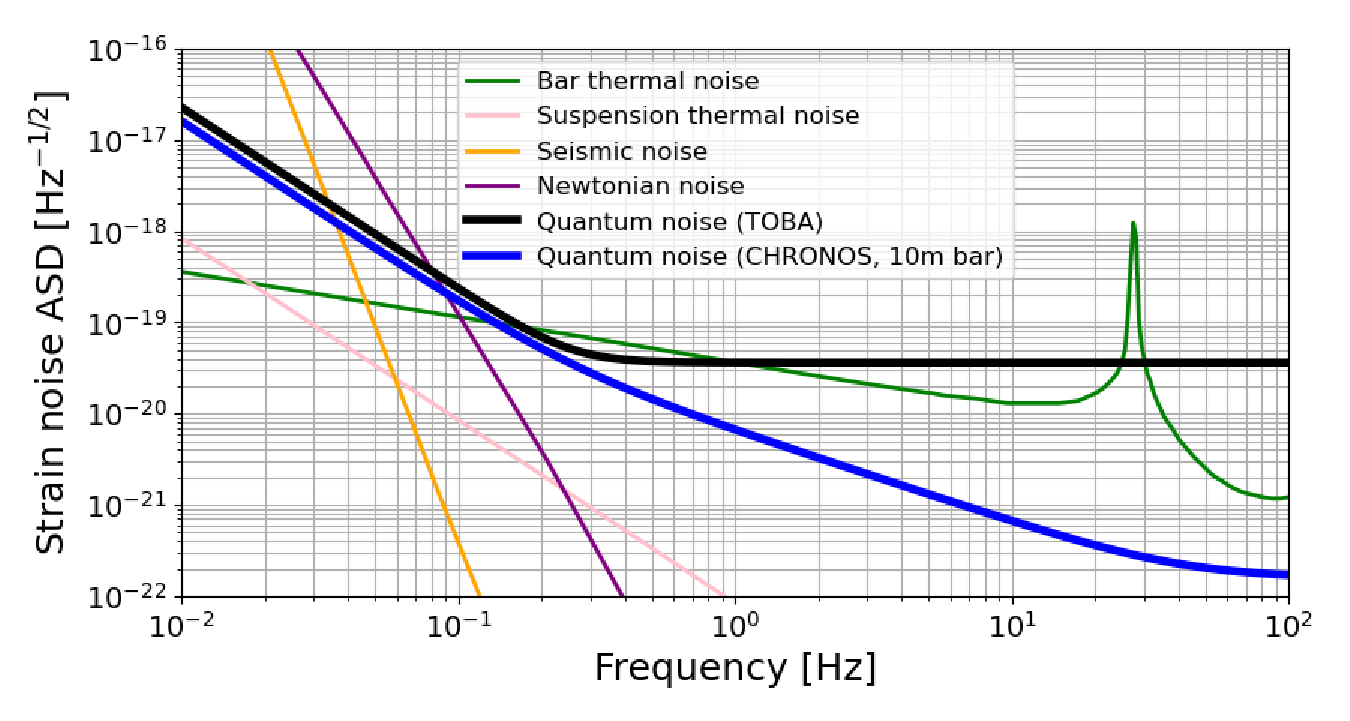}
    \caption{
Comparison of the strain sensitivity between TOBA and CHRONOS.
All non-quantum noise contributions are assumed to be identical to those in Ando \textit{et al.}~\cite{Ando2010}.
The TOBA quantum noise is limited by the standard quantum limit, whereas CHRONOS employs a speed-meter configuration with a 10 m bar length, enabling broadband sensitivity improvement.}
    \label{fig:TOBA}
\end{figure}

We investigated how QND techniques can enhance torsion-bar GW detectors.
As shown in Fig.~\ref{fig:TOBA}, we performed a direct sensitivity comparison assuming the same moment of inertia, mass, bar length, and bar resonance frequency as those of TOBA. All non-quantum noise contributions were taken to be identical to the model used in Ando \textit{et al.} ~\cite{Ando2010}, ensuring that the difference in sensitivity arises solely from the treatment of quantum noise.

The quantum noise of TOBA is constrained by the SQL, leading to a characteristic knee frequency determined by the trade-off between shot noise and radiation-pressure back-action noise. Improving the high-frequency sensitivity of TOBA requires increasing the laser power to suppress shot noise; however, this unavoidably enhances radiation-pressure noise, thereby degrading the low-frequency sensitivity.
In contrast, CHRONOS adopts a speed-meter configuration, which enables the suppression of shot noise and bar thermal noise while maintaining radiation-pressure noise at a comparable level in the low-frequency band. In this analysis, the phase parameters are set to $\phi_p = 40^\circ$, $\phi_s = 0^\circ$, and $\zeta = -70^\circ$. By measuring the velocity (or momentum) of the torsion bars rather than their position, the speed-meter configuration effectively circumvents the SQL-imposed trade-off, resulting in broadband sensitivity improvement, particularly toward lower frequencies. While realizing 10 m-scale sapphire or silicon torsion bars presents significant technical challenges in fabrication, suspension, and cryogenic operation, the combination of torsion-bar detectors with QND speed-meter techniques offers a promising pathway to unlock performance beyond that of conventional TOBA designs. However, even if there is 1m torsion bar design, CHRONOS have enough science reaches.

The test masses are assumed to be sapphire mirrors with a diameter of $22~\mathrm{cm}$, identical to those used in KAGRA~\cite{Ushiba:2019kzq}, and connected by hydro-catalysis bonding~\cite{Elliffe:2005}. Cryogenic operation plays a crucial role in this design: it relaxes the mechanical $Q$ requirements of the bonding interface and suppresses thermal lensing without relying on extreme laser power or high finesse. These features make the implementation of QND techniques in torsion-bar detectors more feasible than in room-temperature, high-power interferometric configurations.

In the future, increasingly precise measurements from cosmic microwave background (CMB) experiments~\cite{Abitbol:2025_SimonsObservatory_LAT, LiteBIRD:2022cnt}, pulsar timing arrays~\cite{Dewdney2009SKA}, and next-generation interferometers~\cite{Adhikari2020VoyagerCryoSi, Punturo2010ET, Evans2021CEHorizon} will provide probes of GW sources across a wide frequency range. In this multi-messenger landscape, CHRONOS measurements will play an important role by directly accessing the sub-Hz regime that remains challenging for other approaches.
\section{Conclusion}
We have presented the design concept of CHRONOS, a next-generation ground-based GW detector optimized for the unexplored $0.1$--$10\,\mathrm{Hz}$ band. 
By combining a triangular Sagnac interferometer with torsion-bar test masses, CHRONOS expects to achieves the first QND measurement of angular momentum in a macroscopic system~\cite{Chen2003,Danilishin2012}. This speed-meter readout suppresses quantum radiation-pressure noise, enabling sub-Hz sensitivity.
With a target sensitivity of $h \simeq5\times10^{-19}\,\mathrm{Hz^{-1/2}}$ at $f\sim2\,\mathrm{Hz}$, CHRONOS will open a new window for GW astronomy. 
It enables detection of intermediate-mass black hole binaries~\cite{Sesana2016}, probing SGWB from the early universe~\cite{Campeti2021Measuring,LiteBIRD:2022cnt}, and detecting prompt gravity signals from large earthquakes~\cite{Harms2015,Montagner2016,Vallee2017}. 
By bridging the gap between LISA~\cite{LISA2017} and future detectors such as Voyager and the Einstein Telescope~\cite{Voyager2020,ET2010}, CHRONOS will establish a crucial platform for multi-band GW and multi-messenger astronomy.

\begin{acknowledgments}
We thank S.Takano, M.Ando and  T.Namikawa for providing the TOBA data and CMB data. 
We also thank K.W.Ng and M.Hazumi for their academic advice during the preparation of this manuscript.
We thank M.Hasegawa, T.Kanayama, S.Matsushita, H.Murakami, M.Inoue, and R.Shibuya for support to establish CHRONOS team. 
Y.I. and H.W. acknowledges support from NSTC, CHiP and Academia Sinica in Taiwan under Grant  No.114-2112-M-008-006-, and No.AS-TP-112-M01.  

\end{acknowledgments}

\bibliographystyle{apsrev4-2}
\bibliography{CHRONOS_PRL}

\end{document}